# Particularities of Analog FCS Optimization

Anatoliy Platonov, *Senior Member, IEEE*

***Abstract*** **– There is analyzed a performance of optimal feedback communication systems with the analog transmitters in the forward channel (AFCS). It is shown that the measures and limit boundaries of AFCS performance are similar but differ from those used in digital communications and information theory. The causes of the differences are discussed.**

## 1. Introduction

The recently used basic criterions of the communication systems (CS) performance are bit error rate (BER), B- (bandwidth or spectral) efficiency $R/F_0$ and P- (power or energy) efficiency of transmission $E^{bit}=W/R$, (e.g.[1]), where $R$ is the bit rate of transmission [bit/s], $E^{bit}$ is the "energy of bit" [J/bit]; $W$ is the power of signals at the channel output, and $F_0$ is the channel bandwidth. The P-B performance of CS is assessed by the closeness of the points ( $R/F_0$ , $E^{bit}=W/R$ ) to the Shannon boundary:

$$\frac{E^{bit}}{N_\xi}=\frac{F_0}{C}\left(2^{\frac{C}{F_0}}-1\right)=\frac{Q^2}{\log_2(1+Q^2)}\quad,\qquad(1)$$

determined for the channels with additive Gaussian noise (AWGN). Value $N_\xi$ in (1) is the double sided spectral power density of AWGN, $Q^2=W/\sigma_\xi^2$ and $\sigma_\xi^2=N_\xi F_0$ are the SNR and power of the noise at the channel output, respectively. Boundary (1) describes "ideal" trade-off between the limit values of P- and B- efficiencies of CS which directly follows from Shannon's formula for the channel capacity:

$$C=F_0\log_2\left(1+\frac{W}{\sigma_\xi^2}\right)=F_0\log_2\left(1+Q^2\right)\quad,\qquad(2)$$

after transition to the variables $R/F_0$ , $E^{bit}/N_\xi$ .

Development of analytical tools for systematic design of the "ideal" CS transmitting signals with bit-rate and P-B efficiencies achieving boundaries (1), (2), remains open task. In [2],[3] we shown that this task can be solved for FCS with the *analog* forward transmission (analog FCS - AFCS). This class of systems allows formulation of the mean square error (MSE) of transmission and further optimization using methods of Bayesian estimation theory [4],[5]. Solution of this task - optimal transmission/reception algorithms [2] enables designing the ideal AFCS. Analysis of the algorithms shown appearance of a series of earlier not studied effects common for AFCS whose performance attains or is close to boundaries (1),(2). The most important of them is the "threshold effect" [2],[3], that is fast aggravation of the limit characteristics of AFCS as a whole (i.e. considered as a generalized communication channel), if the samples of the origin signal are transmitted longer than definite interval of time.

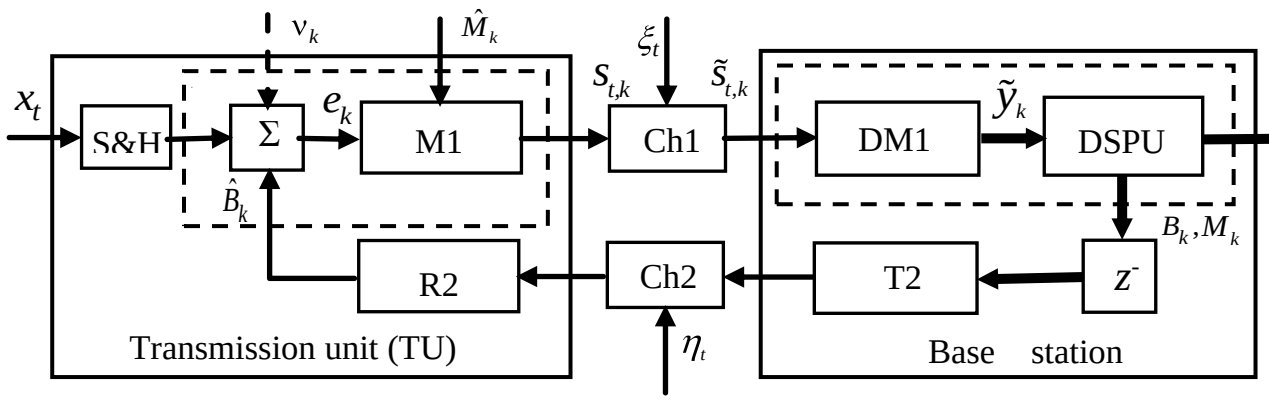

Figure 1. Block-diagram of the analog FCS (AFCS).

Optimal analog FCS were a subject of intensive researches in 1950-1960 ([6]-[8] and other works) and are considered now as a "passé" stage of the CS theory development. However, results presented below and in [2],[3], [5] show that analog CS need development of independent mathematical tools enabling their optimization and solving the tasks unsolvable in the frame of the digital theory.

## 2. AFCS description

The considered systems (Fig. 1) include the transmitting unit (TU) and base station (BS) connected by the forward M1-Ch1-DM1 and feedback T2-Ch2-R2 channels. There is assumed that both channels Ch1, Ch2 are linear, memoryless, and noises $\xi_t$ , $\eta_n$ are AWGN with known spectral power densities. The input signals $x_t$ are stationary Gaussian processes with known mean $x_0$ and variance $\sigma_0^2$ .

### A. Communication scheme

The input signal $x_t$ is sampled in the sample-and-hold unit (S&H). Each sample $x^{(m)}=x(mT)$ ( $m=1,2,...$ ; $T=1/2F$ is sampling period) is routed to the first input of the subtracting unit $\Sigma$ and transmitted in $n=T/\Delta t=F_0/F$ cycles, each of duration $\Delta t=1/2F_0$ , where $2F_0$ is the channels bandwidth. We assume that each sample is transmitted independently and in the same way that permits to reduce the analysis of AFCS functioning to the single sample transmission (for this reason, index $m$ in $x^{(m)}$ is further omitted).

Adaptive modulator includes the subtracting unit $\Sigma$ and amplitude modulator M1 with adjustable modulation index $\hat{M}_k$ . At the beginning of each $k$-th cycle of transmission $(k=1,...,n)$ , $\hat{M}_k$ is set to a definite, additionally determined value. Simultaneously, signal $\hat{B}_k$ at the second input of subtractor $\Sigma$ is set to the value computed by digital processing unit (DSPU) of BS in the previous cycle and delivered to TU over the feedback channel. In this case, the dependence between the values $\hat{B}_k$ and the controls $B_{k-1}=B(\tilde{y}_1^{k-1})$ formed by DSPU can be written as follows:

$$\hat{B}_k=B_{k-1}+\nu_k=B(\tilde{y}_1^{k-1})+\nu_k,\qquad(3)$$

where $\tilde{y}_1^{k-1}\triangleq(\tilde{y}_1,...,\tilde{y}_{k-1})$ are the sequences of signals (observations) formed by the demodulator DM1 in previous cycles of the sample transmission. Variable $\nu_k$ in (3) describes the errors of feedback transmission which can be considered as AWGN with the variance $\sigma_v^2$ dependent on the chosen method of transmission and characteristics of the channel. The assumed in (3) case of one-cycle delay can be generalized in further investigations.

Formed by the subtractor, difference signal $e_k = x - \hat{B}_k$ is modulated and transmitted to BS. The signal $\tilde{s}_{t,k}$ at the output of the channel Ch1 is described by the relationship:

$$\tilde{s}_{k,t} = \frac{\gamma_0}{r} s_{k,t} + \xi_t \ , \qquad (4)$$

where $\gamma_0$ is the channel gain and $r$ is the distance between TU and BS. Observation $\tilde{y}_k$ formed by demodulator DM1 is routed to the input of DSPU. New estimate $\hat{x}_k$ of the sample is computed according to the Kalman-type equation:

$$\hat{x}_k = \hat{x}_{k-1} + L_k[\tilde{y}_k - E(\tilde{y}_k \mid \tilde{y}_1^{k-1})] \ ; \ \ (\hat{x}_0 = x_0) \ , \qquad (5)$$

where $E(\tilde{y}_k \mid \tilde{y}_1^{k-1})$ is one-step prediction of the signal at the demodulator output, and parameter $L_k$ determines the convergence rate of algorithm (5). Computation of $B_k = B(\tilde{y}_1^k)$ and its transmission to TU finishes the cycle. The synchronizing units switch the parameters $\hat{M}_k$ , $\hat{B}_k$ , $L_k$ , to the values $\hat{M}_{k+1}$, $\hat{B}_{k+1}$, $L_{k+1}$, and the next cycle of transmission begins. After $n$ cycles, final estimate $\hat{x}_n$ is routed to addressee, and AFCS begins transmission of the next sample.

### B. *Solution of saturation errors problem*

Analysis of [6]-[8] and other works on AFCS optimization, has shown that difficulties in development of AFCS theory were caused by application of a *linear* model of the modulator (transmitter) M1. This made impossible consideration of the abnormal errors caused by possible saturation of the transmitter and crucially worsening performance of the systems. In our researches this difficulty is removed by application of the saturation-type model:

$$s_{t,k} = A_0 \begin{cases} \hat{M}_k (x - \hat{B}_k) & \text{if} \ \ \hat{M}_k \, | x - \hat{B}_k | \le 1 \\ \text{sign} \ (x - \hat{B}_k) & \text{if} \ \ \hat{M}_k \, | x - \hat{B}_k | > 1 \end{cases} \cos(2\pi f_0 t + \varphi_k) \qquad (6)$$

The values $A_0$, $f_0$, $\varphi_k$ in (6) are the parameters of the carrier, and $(k\text{-}1)\Delta t_0 \le t \le k\Delta t_0$ . To simplify the form of analytical results, the DSB-SC AM is considered, but the obtained results can be easily extended to other types of AM.

Like in [2], [4]-[8], the basic criterion of AFCS performance is MSE of estimates $P_k = E[(x - \hat{x}_k)^2]$ , $(k = 1,...,n)$. Optimization of the system consists in definition of the parameters $\hat{M}_k$ , $L_k$ and controls $B_{k-1} = B(\tilde{y}_1^{k-1})$ which minimize MSE $P_k$ for each $k = 1,...,n$.

Nonlinear model (6) makes direct solution of the task impossible. However, it can be accurately solved using small parameter naturally introduced by the *statistical fitting condition* [4],[5]. This condition is formulated as a constraint imposed, for each $k = 1,...,n$, on the probability of saturation

$$\Pr_k^{lin} = \Pr(\hat{M}_k \, | x - \hat{B}_k | \le 1 \big| \, \tilde{y}_1^{k-1}, B_0^{k-1}, \, M_0^{k-1}) =$$
$$= \Pr(\hat{M}_k \, | x - \hat{B}_k | \le 1 \big| \, \tilde{y}_1^{k-1}) \ge 1 - \mu \qquad (7)$$

which determines the set of "permissible" values of adjustable parameters $\hat{M}_k$ , $\hat{B}_k$ excluding appearance of saturation with a probability not smaller then 1-$\mu$ , $\mu << 1$. Condition (7) guarantees that overwhelming majority of the samples will be transmitted in the linear mode. This permits to replace model (6) by the linear one and to present the signal at the demodulator M1 output in the form:

$$\tilde{y}_k = A \, \hat{M}_k \, (x - \hat{B}_k) + \xi_k \, , \ \text{where} \ \ A = A_0 \, \frac{\gamma_0}{r\sqrt{2}} \, , \qquad (8)$$

$\xi_k$ is AWGN with the variance $\sigma_\xi^2 = N_\xi F_0$ , and $\sqrt{2}$ appears due to demodulation. In this case, one can easily find the (conditional) extreme of MSE of transmission. Cases of saturation may add to the result only values of $O(\mu)$ order.

### C. *Optimal transmission-reception algorithm* [2]

a) For each $k = 1,...,n$ , parameters of the analog modulator (6) should be set to the (optimal) values:

$$\hat{B}_k = \hat{x}_{k-1}(\tilde{y}_1^{k-1}) + \nu_k \ ; \ \ \hat{M}_{k-1} = \frac{1}{\alpha\sqrt{\sigma_v^2 + P_{k-1}}} \qquad (9)$$

and $\hat{x}_0 = x_0$; $\hat{M}_0 = (\alpha\sigma_0)^{-1}$. The values $\hat{M}_{k-1}$ in (9) do not depend on observations, which allows presetting the modulation index to these values. The controls $B_{k-1} = E(x \mid \tilde{y}_1^{k-1}) = \hat{x}_{k-1}(\tilde{y}_1^{k-1})$ transmitted to TU are one-step predictions of the input signal value. Parameter $\alpha$ (saturation factor) is connected with the permissible probability of TU saturation by the equation $\mu = 1\text{-}2\Phi(\alpha)$, where $\Phi(\alpha)$ is the Gaussian error function.

b) Digital unit of BS computes the estimates of the sample according to the equation:

$$\hat{x}_k = \hat{x}_{k-1} + L_k \tilde{y}_k \ ; \ \ \ (\hat{x}_0 = x_0 \ ), \qquad (10)$$

where gains $L_k$ are set, in each cycle, to the values

$$L_k = \frac{1}{AM_{k-1}}(1 - P_k \, P_{k-1}^{-1}) = \frac{1}{AM_{k-1}} \frac{Q^2}{1+Q^2} \frac{P_{k-1}}{\sigma_v^2 + P_{k-1}} \ . \qquad (11)$$

c). Corresponding minimal values of MSE (MMSE) of estimates satisfy the equation:

$$P_k = \frac{(\sigma_\xi^2 + A^2 M_{k-1}^2 \sigma_v^2) P_{k-1}}{\sigma_\xi^2 + A^2 M_{k-1}^2 (\sigma_v^2 + P_{k-1})} =$$
$$= (1+Q^2)^{-1} \left[ \frac{(1+Q^2)\sigma_v^2 + P_{k-1}}{\sigma_v^2 + P_{k-1}} \right] P_{k-1} \ ; \ \ (P_0 = \sigma_0^2) \ . \qquad (12)$$

Parameter $Q^2$ in (11), (12) describes the signal-to-noise ratio (SNR) at the forward channel M1-Ch1-DM1 output:

$$Q^2 = \frac{W}{\sigma_\xi^2} = \frac{A^2 M_{k-1}^2 E[(x - \hat{B}_k)^2]}{\sigma_\xi^2} = \frac{1}{N_\xi F_0} \left( \frac{A_0 \gamma_0}{\alpha r} \right)^2 \ , \qquad (13)$$

where $W = (A/\alpha)^2$ is the power of information component of demodulated signal, constant for each $k = 1,...,n$.

Algorithm (6),(8)-(12) contains the basic information permitting to design AFCS transmitting the signals with minimal MSE. Its particularity is dependence of MSE on the channel bandwidth $F_0$ only through the SNR $Q^2$ and on the baseband of the input signals hidden in the number of the transmissions cycles: $F_n = F_0 \ / \ n$. Thereby, designing of optimal AFCS can be organized in two ways: under given $F_0$ or $F$ . If the bandwidth $F_0$ , power of the transmitter and characteristics of the channels are given, SNR $Q^2 = W \ / \ N_\xi F_0$ is constant. Then MMSE (12) and parameters (9), (10) depend only on the number of transmission cycles and can be easily computed for each $k = 1,...,n$. If the baseband $F$ is given, then SNR $Q^2 = W \ / \ nN_\xi F$ depends on $n$ that radically complicates computation of MMSE and parameters (9),(10). This means that more preferable way of theoretical researches and design of the optimal AFCS is consideration of the bandwidth $F_0$ as a given parameter of the project.

## 3. LIMIT CHARACTERISTICS OF AFCS PERFORMANCE

### *A. Lower boundary of MSE* [2],[3]

Under natural assumption that SNR at the input and output of the forward channel are related by the inequality:

$$SNR_{inp}^{Ch1} = \frac{\sigma_0^2}{\sigma_v^2} >> 1 + Q^2 = 1 + SNR_{out}^{Ch1} \ , \qquad (14)$$

(input signals are pre-amplified, and power of BS transmitter provides sufficiently good feedback channel with small $\sigma_v^2$), MMSE (12) can be represented by the approximate formula:

$$P_n = \begin{cases} \sigma_0^2 (1+Q^2)^{-n} & \text{for } 1 \le n \le n^* \\ \sigma_v^2 (n - n^* + 1)^{-1} & \text{for } n > n^* \end{cases} , \qquad (15)$$

where threshold point $n^*$ is the solution of the equation $P_{n^*} = \sigma_v^2$ and has the form:

$$n^* = \frac{1}{\log_2(1+Q^2)} \log_2\left(\frac{\sigma_0^2}{\sigma_v^2}\right) = \frac{\log_2(SNR_{inp}^{Ch1})}{\log_2(1+SNR_{out}^{Ch1})}. \qquad (16)$$

The obtained results allow us to conclude:

*Claim 1. Common particularity of optimal AFCS realized according to algorithm* (6),(8)-(12) *is the threshold effect: in the interval* $1 \le n \le n^*$*, MMSE of transmission diminishes exponentially, and for* $n \ge n^*$*, with the hyperbolical rate.*

The reason for the effect is that in the interval $1 \le n \le n^*$ optimal adjusting the modulator suppresses influence of the noise $\xi_n$ on the estimates $\hat{x}_k$ of the sample. For $n \ge n^*$, forward channel works as a practically noiseless channel transmitting the residuals $e_k = x - \hat{x}_{k-1} + v_k$. Slow decrease of MMSE is provided by digital processing a weak (power less than $\sigma_v^2$) informative component $x - \hat{x}_{k-1}$ in the noise $v_k$.

### *B. Information limits for the forward channel*

From formulas (8),(9), it follows that $E(\tilde{y}_k \mid \tilde{y}_1^{n-1}) = 0$ and $E(\tilde{y}_k^2 \mid \tilde{y}_1^{n-1}) = E(\tilde{y}_k^2) = (A/\alpha)^2 + \sigma_\xi^2$. The latter means that:

*Claim 2. The signals transmitted by optimal AFCS through the forward channel are AWGN. The mean power of emitted signals has constant value* $W_0 = (A_0/\alpha)^2/2$, *maximal under given probability of saturation* $\mu$.

The prior and posterior entropies of the sequences $\tilde{y}_1^n$,and mutual amount of information in $\tilde{y}_1^n$ and $e_1^n$ take the values:

$$H(\tilde{Y}_1^n) = \sum_{k=1}^{n} H(\tilde{Y}_k) = \frac{n}{2}\log_2\left[2\pi e\sigma_\xi^2\left(1+\frac{A^2}{\alpha^2\sigma_\xi^2}\right)\right] , \qquad (17)$$

$$H(\tilde{Y}_1^n \mid e_1^n) = \frac{n}{2}\log_2(2\pi e\sigma_\xi^2) , \qquad (18)$$

$$I(\tilde{Y}_1^n; e_1^n) = n[H(\tilde{Y}) - H(\tilde{Y} \mid e)] = \frac{n}{2}\log_2(1+Q^2) . \qquad (19)$$

Under given mean value $x_0$ and variance $\sigma_0^2$, formula (19) determines [9] maximal amount of information delivered to BS by the received sequences $\tilde{y}_1^n$. The entropies $H(\tilde{Y}_1^n)$ are additionally maximized over the mean power of the emitted signals. For this reason, one may claim that:

*Claim 3. Optimal AFCS transmits the signals through the forward channel with bit rate equal to its capacity:*

$$R_n^{Ch1} = I_{\max}(\tilde{Y}_1^n; e_1^n)/n\Delta t = F_0 \log_2(1+Q^2) = C = \qquad (20)$$
$$= F^{Ch1}\log_2\left(1+\frac{W}{N_\xi F^{Ch1}}\right) = F^{Ch1}\log_2\left[1+\frac{1}{N_\xi F^{Ch1}}\left(\frac{A_0\gamma_0}{\alpha r}\right)^2\right].$$

*independently from duration of the samples transmission.*

Formula (20) has the same form as (2). The single difference is that (20) directly depends on the saturation factor $\alpha = \Phi^{-1}[(1-\mu)/2]$. In turn, $n\mu$ determines the mean percent of the samples distorted by saturation, i.e. the mean percent o0f erroneous bit-words (word error rate WER) or erroneous bits (BER) in the sequences delivered to the addressee [2].

*Claim 4. Formula* (20) *establishes dependence between the capacity of the forward channel and BER of transmission.*

Simultaneously, relationship (20) confirms known result [10]: capacity of the forward channel does not depend on the characteristics of Gaussian feedback channel.

### *C. Rate distortion function and capacity of AFCS*

The sequences of Gaussian samples generated by the source and independently transmitted, are AWGN with the power $\sigma_0^2$. The prior and posterior entropies of the samples are $H(X) = 1/2 \cdot \log_2(2\pi e\sigma_0^2)$, $H(X \mid \hat{X}_n) = 1/2 \cdot \log_2(2\pi e P_n)$, respectively. Each sample is transmitted in *n* cycles that is in $T_n = n\Delta t = n/2F_0$ [s]. Then, minimal mean number of bits per second permitting to restore the origin signal with tolerance $P_n$ (rate distortion) is determined by the relationship ([11], Theorem 22):

$$R_n^{AS} = \frac{I(X;\hat{X}_n)}{T_n} = \frac{H(X) - H(X \mid \hat{X}_n)}{n\Delta t_0} = \frac{F_0}{n}\log_2\frac{\sigma_0^2}{P_n} . \quad (21)$$

On the other hand, MSE of transmission $P_n$ for optimal AFCS realized according to (6),(8)-(12) attains minimal values (12),(15). Then, substitution of (15) in (21) will define the upper boundary of the mean bit rate at the AFCS output, and this value can be considered as the *capacity of AFCS as a whole* under given BER $n\mu$:

$$\bar{R}_n^{AS} = \begin{cases} F_0 \log_2(1+Q^2) & \text{for } 1 \le n \le n^* \\ \dfrac{F_0}{n}\log_2\left[\dfrac{\sigma_0^2}{\sigma_v^2}(n-n^*+1)\right] & \text{for } n > n^* \end{cases} \text{[bit/s]}. \quad (22)$$

The latter allows us to formulate the claim:

*Claim 5. For the optimal AFCS built according to* (6),(8)-(12), *rate distortion function* (21) *determines the capacity of AFCS considered as a generalized communication channel.*

In turn, formula (22) permits to formulate the following important for practice conclusion:

*Claim 6. Capacity of AFCS is constant and equal to the capacity of the forward channel* $C^{AFCS} = F_0 \log_2(1+Q^2)$ *only if the samples are transmitted not longer than in* $T_{n^*} = n^*/2F_0$ *[s]. Smaller than* $2F_0/n^*$ *sampling frequency and transmission of the samples longer than in* $T_{n^*}$ *[s] decreases the capacity of AFCS.*

The cause of the effect is explained at the end of Sect. 3.A.

### *D. Limit power-bandwidth efficiency of AFCS*

The upper boundary of the bandwidth (spectral) efficiency of AFCS ( $R_n^{AS}/F_0$ bps/Hz) is determined by (22). The upper boundary of power efficiency is determined by the following relationship:

$$\frac{E_n^{bit\ AS}}{N_\xi} = \frac{WT_n}{N_\xi I(X,\hat{X}_n)} = \frac{nW}{2N_\xi F_0\, I(X,\hat{X}_n)} = \frac{nQ^2}{\log_2\dfrac{\sigma_0^2}{P_n}} , \qquad (23)$$

that results in

$$\frac{E_n^{bit\ AS}}{N_\xi} = \begin{cases} \dfrac{Q^2}{\log_2(1+Q^2)} = \dfrac{E^{bit}}{N_\xi} & \text{for } 1 \le n \le n^* \ ; \\ \dfrac{nQ^2}{\log_2\left[\dfrac{\sigma_0^2}{\sigma_v^2}(n-n^*+1)\right]} & \text{for } n \ge n^* \ . \end{cases} \quad (24)$$

Formulas (22)- (24) permit to formulate the claim:

*Claim 7. Transmission of the input signals sampled with the frequency* $2F \le 2F_0 / n^*$ (*i.e. under spectrum expansion greater than* $n^* = F_0 / F$ ) *decreases the capacity and limit values of P-B efficiency of AFCS as a whole.*

These formulas also establish the general form of ideal trade-off between the limit P-B efficiencies of AFCS:

$$\frac{R_n^{AS}}{F_0} \cdot \frac{E_n^{bitAS}}{N_\xi} = Q^2 = SNR_{out}^{Ch1} \quad (25)$$

or

$$\frac{R_n^{AS}}{F_0}[dB] + \frac{E_n^{bitAS}}{N_\xi}[dB] = Q^2[dB]. \quad (26)$$

The formulated claims were verified in simulations for the sequences of $M = 5.000$ Gaussian samples processed by algorithm (6), (9)-(12). The empirical values of P-B efficiencies were computed using empirical MSE $\hat{P}_n = \sum_{m=1}^{M}[x^{(m)} - \hat{x}_n^{(m)}]^2 / M$ and formulas (21), (23).

The plots in Fig. 2 show that transmission of the samples longer than in $n^* = F_0 / F$ cycles ($T > T^* = n^* / 2F_0$ [s]) increases the lower boundary of "energy of bit" i.e. worsens P-efficiency. According to (26), this also worsens B- (spectral) efficiency of AFCS /(the lines for $n = 1$ in Figs 2, 3 coincide with the forward channel P-B efficiencies). The plots for AFCS P-B efficiency (Fig. 3) show that, for $n > n^* = F_0 / F$, it declines from Shannon's boundary (1), and its aggravation appears the earlier, the smaller values $n^*$. At the interval $[1, n^*]$, optimal AFCS transmits the samples "ideally" with P-B efficiency (1). Termination of transmission in shorter number of cycles ($n < n^*$) means not efficient utilization of AFCS: MSE $P_n^{fin}$ will be greater than $\sigma_v^2$ and the same value can be obtained using less powerful forward transmitters.

## 4. Discussion of results

The results presented above, also in [2],[4],[5] and other works show that AFCS (and other analog CS) have own criterions of performance and analytical tools similar but different from those used in digital CS theory.

Digital CS are designed for the fastest and reliable delivering to addressee the *numbers of messages* (or blocks of messages) generated by discrete sources. The messages are transmitted in three stages: source, channel codling and modulation (continuous signals are transmitted after previous sampling and quantization). Definition of the distance between the origin and received numbers of messages has no physical sense, except of the bit sequences transmission. The MSE of transmitted analog signals is only "post factum" characteristic of the systems performance inapplicable to their optimization on the sets of possible methods of quantization, coding and modulation. The basis criterions of performance are BER and P-B efficiency (Sect. 1). The rate distortion is a subject of not frequent theoretical researches.

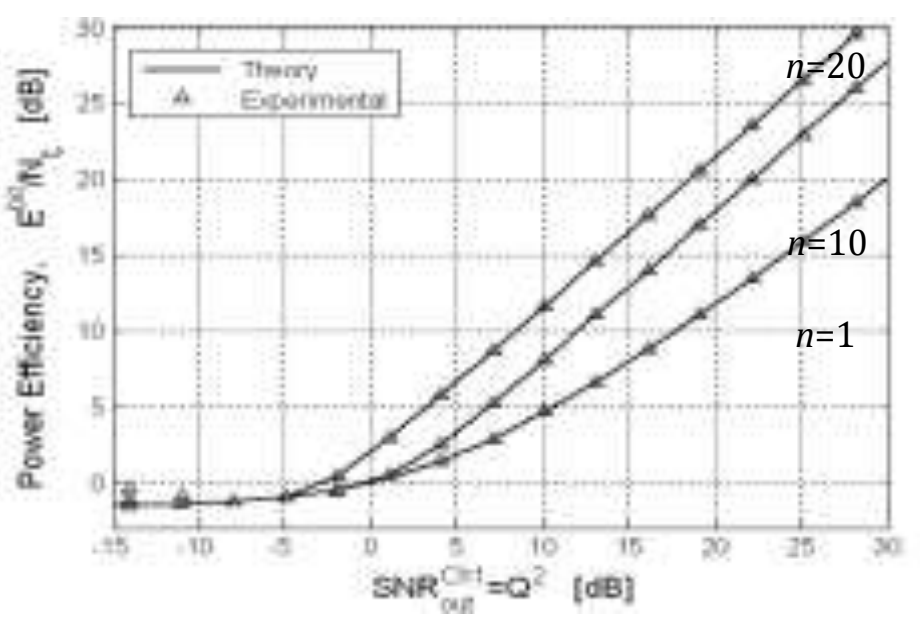

Figure 2. Changes of the limit P-efficiency of AFCS as a function of SNR $Q^2 = W / N_\xi F_0$ after $n = 1;\ 10, 20$ cycles of the sample transmission.

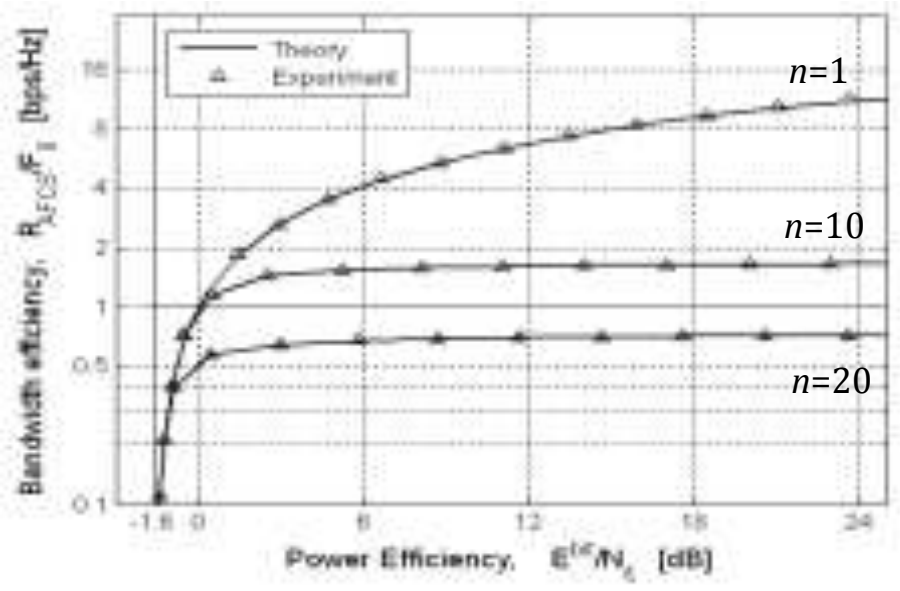

Figure 3. Changes of the limit P-B efficiency of AFCS after $n = 1;\ 10, 20$ cycles of the sample transmission.

In turn, analog transmission employs solely sampling and modulation, and AFCS (and other analog CS) allow direct optimization by the methods of Bayesian estimation theory. The basic performance criterion is MMSE, and adequate instrument of the analog systems analysis and design are the rate distortion and Bayesian estimation theory. Information characteristics of AFCS are derivatives of MSE. Moreover, lose less analog transmission makes the limit characteristics obtained for AFCS valid also for digital FCS transmitting the analog signals in the same conditions.

## References

[1] A Chen *et al.*, Fundamental Tradeoffs on Green Wireless Networks, *IEEE Commun. Mag.*, vol. 49, no. 6, June, 2011, pp. 30–37.

[2] A. Platonov, Capacity and power-bandwidth efficiency of wireless adaptive feedback communication systems, *IEEE Comm. Letters*, vol. 16. no. 5, pp. 573-576, 2012.

[3] A. Platonov, A.Platonov, "Optimization of adaptive communication systems with feedback channels", *IEEE Wireless Comm. and Networking Conf. WCNC'2009*, Budapest, IEEE Xplore, Apr. 2009.

[4] A. Platonov, *Analytical methods of analog-digital adaptive estimation systems design*, D.Sc. monograph, Publishing House of Warsaw Univ. of Technology, s. "Electronics", vol. 154, Warsaw, 2006 (in Polish).

[5] A. Platonov, Optimal identification of regression-type processes under adaptively controlled observation, *IEEE Trans. on Sign. Proc.*, vol. 42, no. 9, Sept. 1994, pp. 2280-2291.

[6] T. Kailath, "An application of Shannon's rate-distortion theory to analog communication over feedback channels", *Proc. IEEE*, vol. 55, no. 6, 1967, pp. 1102-1103.

[7] T. J. Goblick, "Theoretical limitations on the transmission of data from analog sources", *IEEE Trans. on Inf. Theory*, vol.11, no.4, 1965, pp. 558 – 567.

[8] J. P. M. Schalkwijk, L. I. Bluestein, "Transmission of analog waveforms through channels with feedback", *IEEE Trans. on Inf. Theory*, vol. 13, no.4, 1967, pp.617-619.

[9] R.G. Gallager, Information Theory and Reliable Communication, J. Wiley, NY, 1968.

[10] C. E. Shannon, "The zero-error capacity of a noisy channel." *IRE Trans. on Inf. Theory*, vol.2, no.3, 1956, pp. 8-19.

[11] C. E. Shannon, Communication in the presence of noise, *Proceedings of IRE*, vol. 37, no. 1, 1949, pp. 10-21.